\documentstyle[11pt]{article}
\twocolumn 
\def\gsim{\;\rlap{\lower 2.5pt
 \hbox{$\sim$}}\raise 1.5pt\hbox{$>$}\;}
\def\lsim{\;\rlap{\lower 2.5pt
   \hbox{$\sim$}}\raise 1.5pt\hbox{$<$}\;}
\newcommand\beq{\begin{equation}}
\newcommand\eeq{\end{equation}}

\def\la{\lsim}
\def\ga{\gsim}
\title{The First Stars and Quasars}
\date{}
\author{Abraham Loeb\\
Harvard Astronomy Department, 60 Garden Street,
Cambridge, MA 02138}

\begin{document}
\maketitle
\centerline{ABSTRACT}
The first star clusters and quasars resulted directly from the growth of
linear density fluctuations in the early Universe. Since they emerged from
a well-defined set of initial conditions, the interplay between
observational data from the Next Generation Space Telescope (NGST) and
theoretical modeling will advance the current understanding of star and
quasar formation.  The first objects had a substantial impact on the
thermal and chemical state of the rest of the Universe as they reionized
the intergalactic medium and enriched it with metals.

\medskip
\centerline{1. INTRODUCTION}
		
The Universe followed the arrow of time in pedagogical order. It started
with the simplest initial state of almost perfect homogeneity and isotropy
and ended up with systems as complex as intelligent organisms at present
(Fig.  1). So far, standard textbooks are unable to fully follow this
pedagogical track because of the lack of observational data about the
Universe in the redshift interval $5\la z\la 10^3$. The COBE satellite
probed the microwave anisotropies which are remnant from the recombination
epoch at $z\sim 10^3$ (Bennett et al.  1996), and existing telescopes are
limited in their range to redshifts $z\la 5$.  The Next Generation Space
Telescope (NGST) will bridge the above gap in our knowledge and image the
first sources of light that had formed in the Universe. With its
exceptional nJy sensitivity in the 1--3.5$\mu$m infrared regime, NGST is
ideally suited for probing optical-UV emission from sources at redshifts
$\ga 10$, just when popular Cold Dark Matter models for structure formation
predict the first baryonic objects to have collapsed (Fig. 2).

The first sources are a direct consequence of the growth of linear density
fluctuations. As such, they emerge from a well-defined set of initial
conditions and the physics of their formation can in principle be coded to
the form of a precise computer simulation.  The cosmic initial
conditions for star formation are much simpler and better defined than
those in the local interstellar medium; they include a prescription for the
primordial power spectrum of Gaussian density fluctuations, the mean
temperature and density of the gas, the light element abundances according
to Big-Bang nucleosynthesis, and the lack of dynamically-significant
magnetic fields. 

\onecolumn
\begin{figure}[p]
\includegraphics{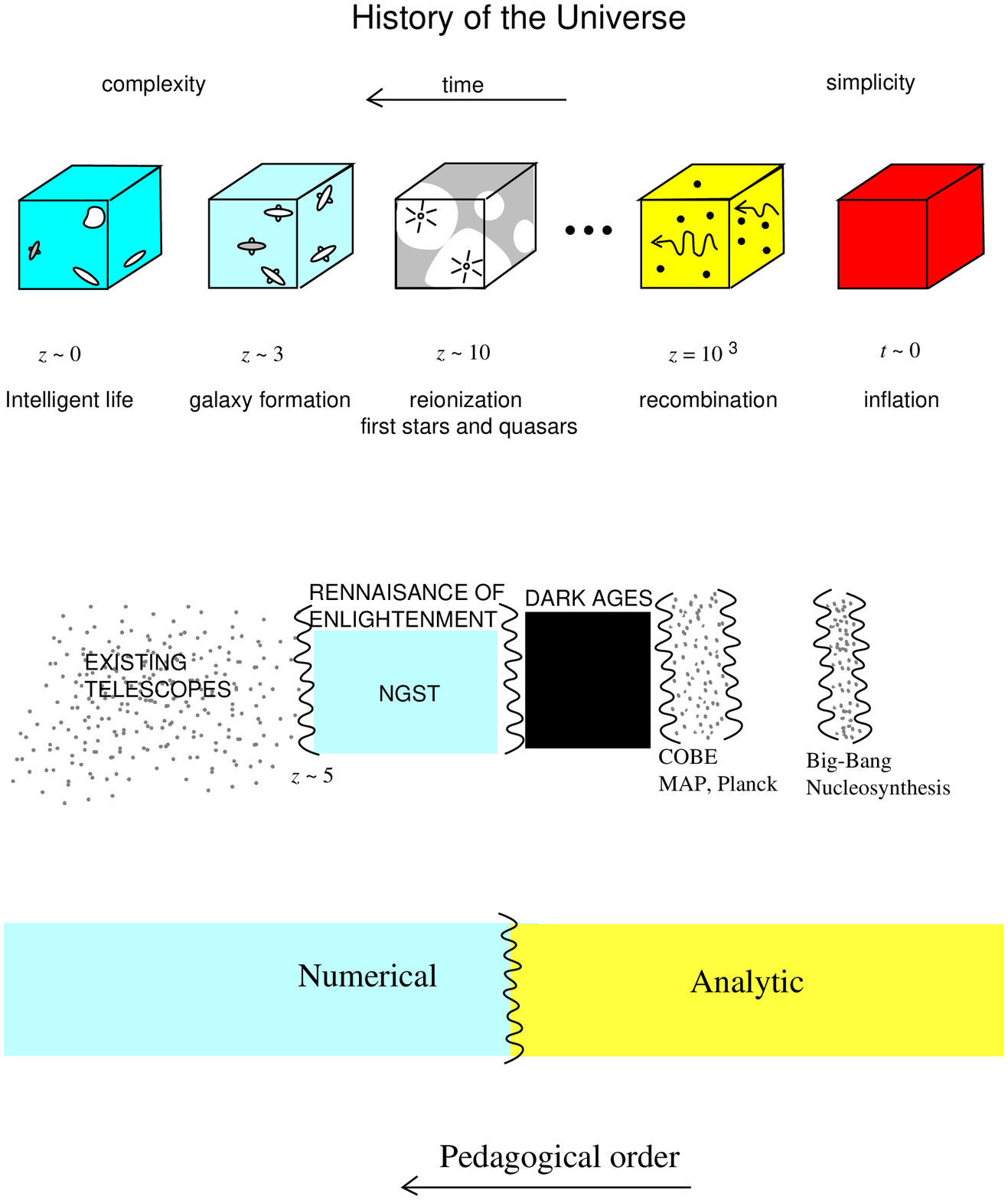}
\vspace{6.2in}
\caption{Milestones in the evolution of the Universe from 
simplicity to complexity. NGST will bridge between the recombination epoch
probed by COBE ($z\sim 10^3$) and the horizon of current observations
($z\sim 5$).}
\label{fig-1}
\end{figure}

\twocolumn
\noindent
The initial mass function of the first stars is determined by this
set of initial conditions (while subsequent stellar generations are
affected by photoionization heating and metal enrichment feedbacks).
Although the early evolution of the seed density fluctuations can be fully
described analytically, the collapse and fragmentation of nonlinear
structure must be simulated numerically.  The first baryonic objects
connect the simple initial state of the Universe to its complex current
state, and their study offers the key to advancing our knowledge on the
formation physics of stars and massive black holes.

\medskip
\centerline{2. THE REIONIZATION EPOCH}
\medskip

The collapse redshifts of baryonic objects in a standard CDM cosmology is
shown in Figure 2. As seen from this figure, more than a tenth of all
baryons could have assembled into virialized objects by a redhift $z\sim
10$.  Since nuclear fusion yields $\sim 7$ MeV per baryon and black hole
accretion might provide even more energy, and since hydrogen ionization
requires 13.6 eV, it follows that even the collapse of only a
negligible fraction ($\ga 10^{-5}$) of all baryons into stars or black
holes could ionize the rest of the Universe. A detailed calculation of the
expected star and quasar populations in the popular $\Lambda$-CDM
cosmology, implies that the Universe is typically reionized at $z\sim 10$
(Haiman \& Loeb 1998a).

\begin{figure}
\includegraphics{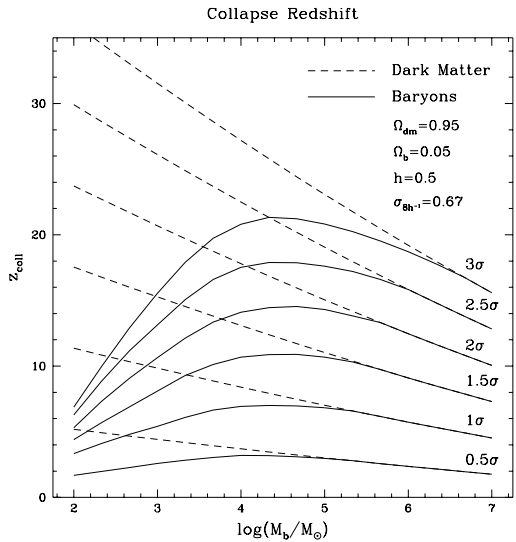}
\vspace{4.5in}
\caption{Collapse redshift, $z_{\rm coll}$, for cold dark matter (dashed lines)
and baryons (solid lines) in spheres of various baryonic masses, $M_{\rm b}$,
and initial overdensities. The overdensities are in units of the rms
amplitude of fluctuations $\sigma(M)$ for a standard CDM power-spectrum
with $\sigma_{8h^{-1}}=0.67$. The collapse of the baryons is delayed
relative to the dark matter due to gas pressure.  The curves were obtained
by following the motion of the baryonic and dark matter shells with a
spherically symmetric, Lagrangian hydrodynamics code (Haiman \& Loeb 1997).
}
\label{fig-2}
\end{figure}

Reionization is defined as the time when the volume filling factor of
ionized hydrogen (HII) approached a value close to unity.  The transition
to this state was 
\onecolumn
\begin{figure}[p]
\includegraphics{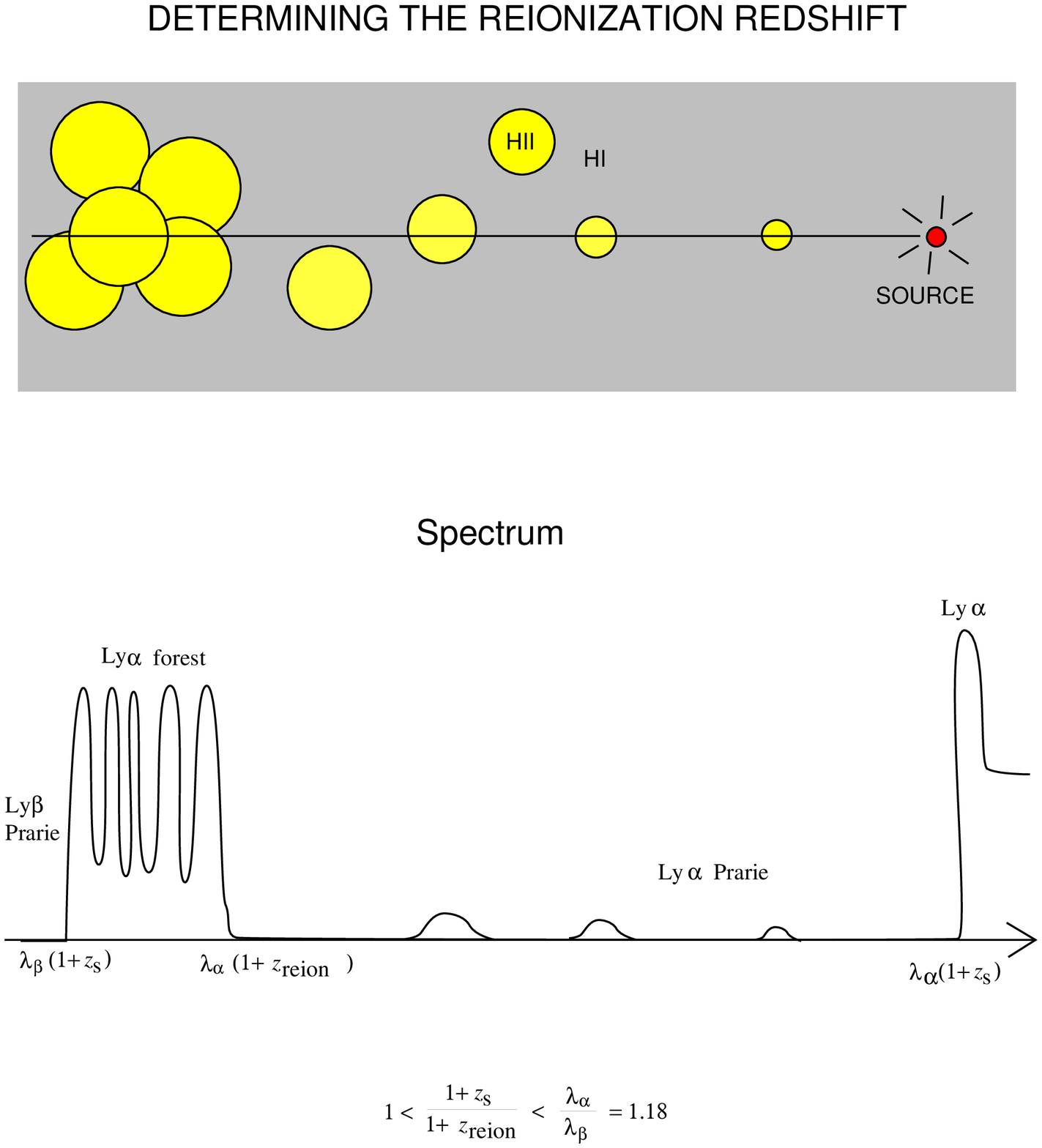}
\vspace{6.2in}
\caption{Sketch of the expected spectrum of  a source at a redshift $z_{\rm s}$
slightly above the reionization redshift $z_{\rm reion}$. The transmitted
fluxes due to HII bubbles in the pre-reionization era and the
Ly$\alpha$ forest in the post-reionization era are exaggerated for
illustration.}
\label{fig-3}
\end{figure}

\twocolumn
\noindent
almost sudden, because the ionizing sources were typically separated by a
distance which is much smaller than the Hubble length so that the HII
regions they had produced overlaped on a timescale much shorter than the
Hubble time. {\it Is it possible to identify the reionization redshift,
$z_{\rm reion}$, from the spectrum of high redshift sources?} The most
distinct feature in the spectrum of a source at $z_{\rm s}> z_{\rm reion}$
would be the Gunn-Peterson absorption trough due to the neutral
intergalactic medium that fills the Universe prior to reionization.  Figure
3 provides a sketch of the spectrum of a source with $(1+ z_{\rm s})>1.18
\times (1+z_{\rm reion})$ which contains some transmitted flux between the
Gunn-Peterson troughs due to Ly$\alpha$ and Ly$\beta$ absorption.  This
transmitted flux is suppressed by the residual Ly$\alpha$ forest in the
post-reionization era.  The possibility of identifying $z_{\rm reion}$ from
the damping wing of the Gunn-Peterson trough (Miralda-Escud\'e 1997)
suffers from potential confusion with damped Ly$\alpha$ absorption along
the line of sight and from ambiguities due to peculiar velocities and the
proximity effect.  An alternative method makes use of the superposition of
the spectra of many sources (Haiman \& Loeb 1998b). In the absence of the
Ly$\alpha$ forest this superposition should result in the sawtooth template
spectrum shown in Figure 4 [with the horizontal axis redshifted by
$(1+z_{\rm reion})$ to the observer's frame]. The cumulative radiation
produced by the first sources might also get reprocessed through
intergalactic dust and appear as a diffuse infrared background (Haiman \&
Loeb 1998a).

The first stars have also enriched the intergalactic medium with metals.
The observed disparity (Lu et al. 1998) between the metallicity of the
Ly$\alpha$ forest ($\la 10^{-2}Z_\odot$) systems and the damped Ly$\alpha$
absorbers ($\sim 0.1 Z_\odot$) indicates that some of the metals are
retained within the potential wells in which they are produced.
Nevertheless, the nonzero metalicity found in the intergalactic gas implies
that winds driven by supernovae, massive stars, gamma-ray bursts (Loeb \&
Perna 1998), or merger events (Gnedin \& Ostriker 1997) expelled some of
the enriched gas out of galaxies.  A typical Type II supernova releases
$\sim 1$ eV in thermal energy per $\sim 10^{-3}Z_\odot$ in metals. The
virial temperature of an object of total mass $M_{\rm tot}$ which formed at
a redshift $z_{\rm f}$ is $\sim 1~{\rm eV} \left(M_{\rm tot}/10^8
M_\odot\right)^{2/3}~\left[(1+z_{\rm f})/10\right]$.  Hence, gas could be
expelled out a potential well only if its metallicity exceeds the minimum
value of $\sim 10^{-3}Z_\odot
\left(M_{\rm tot}/10^8 M_\odot\right)^{2/3}\left[(1+z_f)/10\right]$. Moreover,
in order for mixing to occur accross the Hubble velocity difference between
star clusters, one needs to supply the gas with a kinetic energy of $\sim
1~{\rm eV}\times [(1+z)/10]^3 \left(d/10~{\rm kpc}\right)^2$, where $d$ is
the average distance between neighboring star clusters. Since objects of
mass $M$ collect their mass from a radius $r\sim 10~{\rm kpc}~(M_{\rm
tot}/10^8 M_\odot)^{1/3}[(1+z)/10]^{-1}$ and since at high redshifts $d\gg
r$, considerable mixing will occur only if the metallicity inside the
objects is increased well above the level of $\sim 10^{-3}Z_\odot$.  The
early stars are expected to behave as collisionless particles
\onecolumn
\begin{figure}[p]
\vspace{6in}
\includegraphics{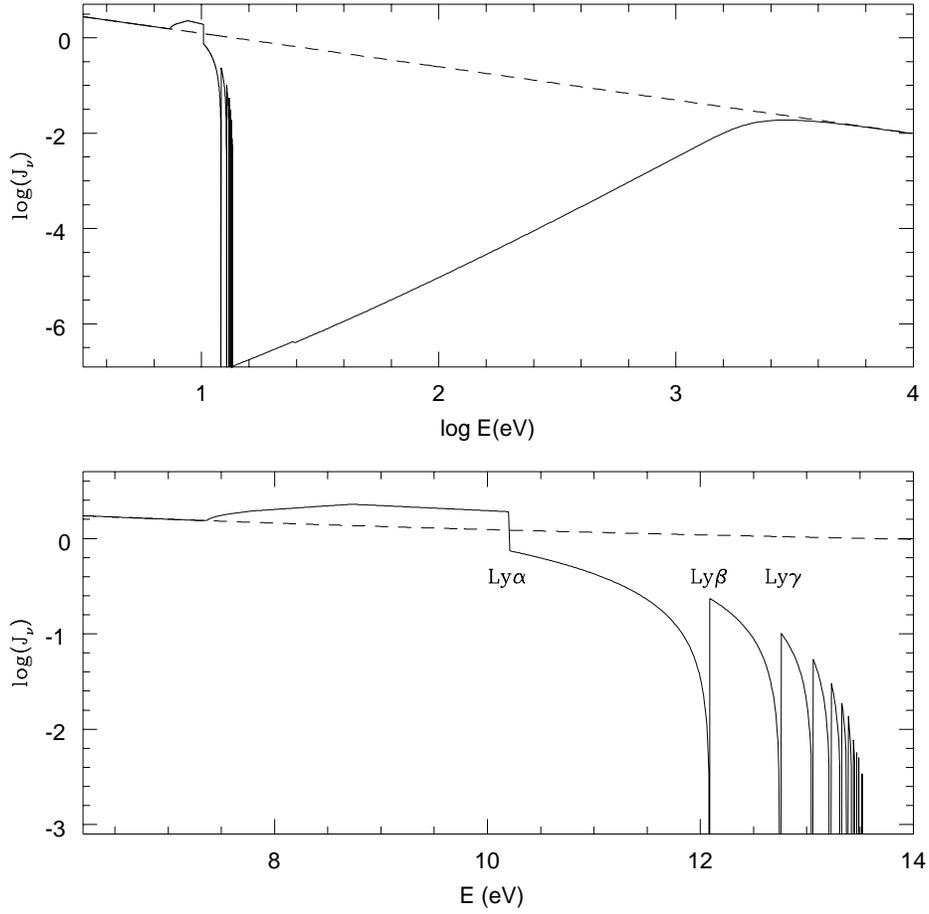}
\caption[Processed Spectrum] {\label{fig4} The combined spectrum 
of many pre-reionization sources, ignoring absorption by the Ly$\alpha$
forest after reionization.  The upper panel shows that absorption by
neutral hydrogen and helium suppresses the flux above 13.6eV up to the keV
range.  The lower panel shows the sawtooth-modulation due to line
absorption below 13.6eV (from Haiman, Rees, \& Loeb 1997). }
\label{fig-4}
\end{figure}

\twocolumn
\noindent
and eventually populate the halos of present-day massive galaxies.
Therefore, direct observations of the abundance patterns in the halo
stellar population of the Milky-Way could uncover the early enrichment
history of the first star clusters and the intergalactic medium.

\centerline{3. DETECTION WITH NGST}

{\it How many sources will NGST see?} Figure 5 shows the predicted number
of quasars and star clusters expected per field of view of NGST (Haiman \&
Loeb 1998b).  In this calculation, the star formation efficiency was
calibrated based on the inferred metallicity range of the Ly$\alpha$ forest
(Songaila \& Cowie 1996; Tytler et al. 1995) while the characteristic
quasar lightcurve was calibrated in Eddington units so as to fit the
observed luminosity function of bright quasars at $z\sim2$--4.  Both
populations of sources were extrapolated to high redshifts and low
luminosities using the Press-Schechter formalism (for more details, see
Haiman \& Loeb 1998a).  Typically, there are of order tens of sources at
redshifts $z>10$ per field of view of NGST. The lack of point source
detection in the Hubble Deep Field is consistent with a low-mass cutoff for
luminous matter in halos with circular velocities $\la 50$--75$~{\rm
km~s^{-1}}$, due to photoionization heating (Haiman, Madau,
\& Loeb 1998).  The redshift of early sources can be easily identified
photometrically based on their Ly$\alpha$ trough.  Figure 5
demonstrates that NGST will play a dominant role in exploring the
reionization epoch and in bridging between the initial and current states
of the Universe.

\begin{figure}
\includegraphics{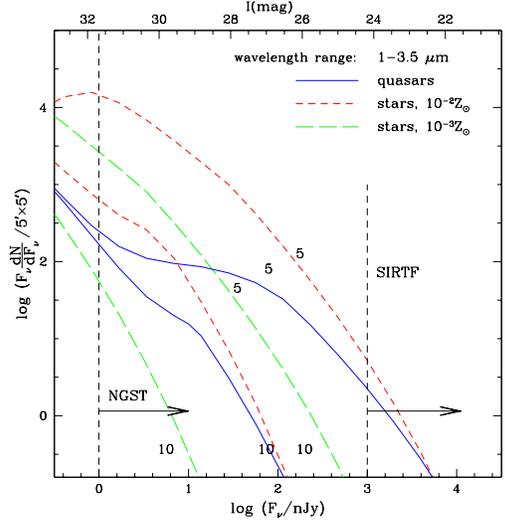}
\vspace{2.8in}
\caption{Predicted number counts per
$5^\prime \times 5^\prime$ field of view per logarithmic flux interval in
the NGST wavelength range of $1$--$3.5\mu$m.  The numbers of quasars and
star clusters were calculated for a $\Lambda$CDM cosmology with
$(\Omega_{\rm M}, \Omega_{\Lambda},
\Omega_{\rm b}, h, \sigma_{8h^{-1}}, n)=(0.35, 0.65, 0.04, 0.65, 0.87,
0.96)$.  The lowest mass scale of virialized baryonic objects was chosen
consistently with the photoionization feedback due to the UV background.
The star formation efficiency was calibrated so as to bracket the possible
values for the average metallicity of the Universe at $z\sim 3$, namely
between $10^{-3}Z_\odot$ and $10^{-2}Z_\odot$.  The thick lines, labeled
``10'', correspond to objects located at redshifts $z>10$, and the thin
lines, labeled ``5'', correspond to objects with $z>5$.  The upper labels
on the horizontal axis correspond to Johnson I magnitude (from Haiman \&
Loeb 1998b).}
\label{fig-5}
\end{figure}

\eject
\centerline{ACKNOWLEDGEMENTS}	

\vskip 0.15in
\noindent
I thank my former student, Zoltan Haiman, for educating me on much of the
material covered by this review, and Sarah Jaffe for assistance with the
artwork.  This work was supported in part by the NASA ATP grant NAG5-3085
and the Harvard Milton fund.

\vskip 0.3in
\centerline{REFERENCES}

\vskip 0.2in
\noindent
Bennett, C. L. et al. 1996, ApJ, 464, L1
\vskip 0.1in
\noindent
Gnedin, N. Y., \& Ostriker, J. P. 1997, ApJ, 486, 581
\vskip 0.1in
\noindent
Haiman, Z., \& Loeb, A. 1997, ApJ, 483, 21
\vskip 0.1in
\noindent
-----------. 1998a, ApJ, in press, astro-ph/9710208
\vskip 0.1in
\noindent
-----------. 1998b, ApJ, to be submitted
\vskip 0.1in
\noindent
Haiman, Z., Madau, P., \& Loeb, A. 1998, ApJ, submitted, astro-ph/9805258
\vskip 0.1in
\noindent
Haiman, Z., Rees, M. J., \& Loeb, A. 1997, ApJ, 476, 458
\vskip 0.1in
\noindent
Loeb, A., \& Perna, R. 1998, ApJL, in press, astro-ph/9805139
\vskip 0.1in
\noindent
Lu, L., Sargent, W. L. W., Barlow, T. A., \& Rauch, M. 1998, AJ, submitted,
astro-ph/9802189
\vskip 0.1in
\noindent
Miralda-Escud\'e J. 1997, ApJ, submitted, astro-ph/9708253
\vskip 0.1in
\noindent
Songaila, A., \& Cowie, L. L. 1996, AJ, 112, 335
\vskip 0.1in
\noindent
Tytler, D. et al. 1995, in QSO Absorption Lines, ESO Astroph. Symposia, ed. G.
Meylan (Heidelberg: Springer), p. 289

\end{document}